\documentclass[aps,prl,reprint,amsmath,amssymb,superscriptaddress]{revtex4-2}

\usepackage{graphicx}
\DeclareRobustCommand{\rchi}{{\mathpalette\irchi\relax}}
\newcommand{\irchi}[2]{\raisebox{\depth}{$#1\chi$}}
\usepackage[usenames,dvipsnames]{xcolor}
\usepackage{dcolumn}
\usepackage{bm}
\usepackage[colorlinks=true,
            allcolors=blue!60!black]{hyperref}

\usepackage{physics}
\usepackage{bbold}
\usepackage{newtxtext} 

\newcommand{\bluecite}[1]{%
	\textcolor{blue!60!black}{{\cite{#1}}}%
}
\newcommand{\rref}[1]{%
	\textcolor{blue!60!black}{({\ref{#1}})}%
}

\begin{document}


\title{Quantification of Quantum Dynamical Properties with Two Experimental Settings}

\author{Tzu-Liang Hsu}
\author{Kuan-Jou Wang}
\author{Chun-Hao Chang}
\author{Sheng-Yan Sun}
\author{Shih-Husan Chen}
\author{Ching-Jui Huang}
\author{Che-Ming Li}
\email{cmli@mail.ncku.edu.tw}
\affiliation{Department of Engineering Science, National Cheng Kung University, Tainan 70101, Taiwan}

\affiliation{Center for Quantum Frontiers of Research and Technology, National Cheng Kung University, Tainan 70101, Taiwan}


\date{\today}

\begin{abstract}
Characterizing quantum dynamics is essential for quantifying arbitrary properties of a quantum process---such as its ability to exhibit quantum-mechanical dynamics or generate entanglement. However, current methods require a number of experimental settings that increases with system size, leading to artifacts from experimental errors. Here, we propose an approximate optimization method that estimates property measures using only two mutually unbiased bases to compute their lower and upper bounds, and to reconstruct the corresponding processes. This system-size independence prevents error accumulation and allows characterization of the intrinsic quantum dynamics. Compared with quantum process tomography, we experimentally validate our method on photonic fusion and controlled-NOT operations, demonstrating accurate resource estimation while substantially reducing the number of required Pauli experimental settings: from 81 to 10 for the photonic fusion and to 2 for the controlled-NOT. These results show that our method is well-suited for estimation of dynamical properties in architectures ranging from chip-scale quantum processors to long-distance quantum networks.
\end{abstract}


\maketitle

\textit{Introduction.}---Quantum evolution governs the transformation of quantum states, yet it inevitably manifests as a noisy quantum process due to unavoidable quantum decoherence and experimental imperfection. Quantifying the properties of quantum processes is crucial for applications ranging from chip-scale quantum computing~\bluecite{chip1,chip2} to long-distance quantum information processing~\bluecite{net1,net2,net3}. To this end, considerable progress has been made in the quantification of dynamical properties. For instance, process fidelity measures the closeness of a quantum process to its intended operation and is widely used to assess process performance~\bluecite{fidelity,Hofmann}. The diamond norm quantifies the distinguishability between quantum channels and plays an important role in quantum process discrimination~\bluecite{dia1,dia2} and quantum resource theory~\bluecite{channelresource1,channelresource2}. Other dynamical measures include process composition and robustness. Composition quantifies how much of quantum processes that can create resource (e.g., coherence or entanglement), while robustness measures how much noise can be added before the process can no longer create any resource~\bluecite{rob1,rob2,quantify,capability}.

\begin{figure}[h!]
    \centering
    \includegraphics{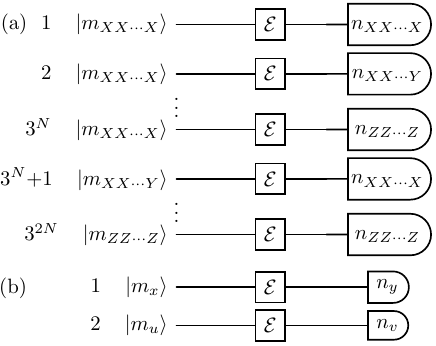}
    \caption{Comparison of experimental settings for SQPT and AOQPT introduced in this work. Each row illustrates an input-output measurement of a quantum process \(\mathcal{E}\) under a single experimental setting.  
		(a) SQPT requires \(3^{2N}\) distinct experimental settings for full \(\rchi\)-matrix reconstruction. Input states are prepared as eigenstates \(\ket{m_P}\) of Pauli operators \(P \in \{X,Y,Z\}^{\otimes N}\), and the corresponding output states are measured in Pauli basis \(\{\ket{n_P}\}\), where \(m,n \in \{0,\ldots,d-1\}\). (b) In contrast, AOQPT requires only two experimental settings, preparing quantum states from a pair of mutually unbiased bases and performing measurements in Eqs.~\rref{eq:Txy} and \rref{eq:Tuv}. The quantum circuits were drawn using Quantikz~\bluecite{Qtikz}.}
    \label{fig:settings}
\end{figure}

In general, an arbitrary property measure \(Q\) of a quantum process \(\mathcal{E}\) can be expressed as
\begin{equation}
    Q = C_Q(\rchi),
    \label{Q}
\end{equation}
where \(\rchi\) is a positive semidefinite superoperator that completely characterizes \(\mathcal{E}\) and can be reconstructed via quantum process tomography~\bluecite{QCQI,chi1,chi2}, and \(C_Q\) maps \(\rchi\) to the property measure of interest. A fundamental method for reconstructing \(\rchi\) is standard quantum process tomography (SQPT). For an \(N\)-qubit process, SQPT involves preparing \(d^2\) separable states and performing quantum state tomography on the corresponding output states to collect \(d^{4}\) real parameters, where \(d=2^N\) is the system dimension. Since these procedures are typically performed in the Pauli bases, both state preparation and measurement each require \(3^N\) settings, leading to \(3^{2N}\) experimental settings. Here, a setting refers to either preparing or projecting onto the eigenstates of a chosen \(N\)-qubit operator, and an experimental setting is a combination of one preparation setting and one measurement setting [see Fig.~\hyperlink{fig:settings}{1(a)}]. See Sec.~I of Supplemental Material (SM) for the detailed discussion of SQPT~\bluecite{SM}. Consequently, this exponential scaling causes evaluating \(Q\) via Eq.~\rref{Q} to be resource-intensive even for small quantum systems~\bluecite{ancilla}.

The goal of this work is to develop a tomographic method for evaluating any \(Q\) in Eq.~\rref{Q} with experimental settings independent of system size. Existing methods~\bluecite{CQPT,VQPT,DCQD,UQPT,BQPT,EAQPT} alleviate resource costs but still scales with increasing dimension. As the system size grows, more distinct experimental settings are needed. The associated state preparation and measurement errors accumulate across them, thereby amplifying artifacts in real-world implementations. Consequently, current methods cannot faithfully capture intrinsic dynamics and hinder reliable estimation of dynamical properties~\bluecite{fidelity,dia1,rob1,capability,quantify}.

To address these issues, we propose a size-independent tomographic method---\textit{approximate optimization quantum process tomography} (AOQPT). Here, approximate optimization refers to estimating dynamical properties by optimizing over a reduced number of experimental settings. We demonstrate that essential information about the deviation of a quantum process from a target operation can be extracted using only two experimental settings---one capturing classical input-output behavior, and the other revealing quantum interference between distinct basis operations. The process property in Eq.~\rref{Q} is then estimated by computing its lower and upper bounds from only \(2d^2\) probabilities, while also reconstructing the worst- and optimal-case processes. This yields an asymptotic reduction by \(1/d^2\) compared to SQPT. Experimentally, we validate AOQPT on a photonic fusion operation~\bluecite{fusion1,fusion2,fusion3,fusion4,fusion5,senior} realized on an optical platform and a controlled-NOT (CNOT) gate implemented on the \texttt{ibmq\_belem} device. Compared to SQPT, the number of required Pauli experimental settings is reduced from 81 to 10 for the photonic fusion and to 2 for the CNOT.

\textit{AOQPT framework.}---Consider an experimental quantum process \(\mathcal{E}\) that aims to realize a target unitary operation \(S\). Let \(\ket{m_y} = S \ket{m_x}\) be the ideal output state with input state \(\ket{m_x}\), where \(m \in \{0, \ldots, d - 1\}\). Given a fixed experimental setting \((x, y)\) of preparation and measurement, the probability that the output of \(\mathcal{E}\), acting on the input state \(\ket{m_x}\), is measured to be in the state \(\ket{n_y}\) is given by the Born’s rule:
\begin{align}
	\nonumber 
	T^{x \rightarrow y}_{mn} &= \bra{n_y} \mathcal{E}(\ketbra{m_x}) \ket{n_y} \\
	&= P(n_y | m_x),
	\label{eq:Txy}
\end{align}
where the diagonal elements \(T^{x \rightarrow y}_{mm}\) represent the probabilities of correct outputs, while the off-diagonal elements \(T^{x \rightarrow y}_{mn}\) with \(m \neq n\) are the probabilities of incorrect outcomes. The second equality shows that \(T^{x \rightarrow y}_{mn}\) are also elements of a stochastic matrix \(T^{x \rightarrow y}\), which describes a classical stochastic process transitioning input \(m_x\) to output \(n_y\)~\bluecite{CIF}. Hence, for any fixed \((x, y)\), the observable dynamics of \(\mathcal{E}\) can be simulated by a classical stochastic process.

To examine the limitations and capabilities of $T^{x \rightarrow y}$ in simulating $\mathcal{E}$, it is useful to express $\mathcal{E}$ as~\bluecite{QCQI}:
\begin{equation}
\mathcal{E}(\rho) = \sum_{p,q,r,s=0}^{d-1}\rchi_{pq,rs} E_{pq}\rho E_{rs}^\dagger,
\label{eq:qpt}
\end{equation}
where the operation elements $E_{pq} = \ketbra{q_y}{p_x}$ form an orthogonal operator basis~\bluecite{orthogonality}. Inserting Eq.~\rref{eq:qpt} into Eq.~\rref{eq:Txy} gives
\begin{equation} 
T^{x \rightarrow y}_{mn} = \rchi_{mn,mn}. 
\label{eq:Txy_chi}
\end{equation} 
It reveals that $T^{x \rightarrow y}$ contains only the diagonal elements of $\rchi$ and thus describes solely an incoherent process, given by~\bluecite{coh}:
\begin{equation}
\mathcal{E}_\text{I}(\rho)=\sum_{p,q=0}^{d-1}\rchi_{pq,pq}E_{pq}\rho E_{pq}^\dagger.
\label{eq:incoherent_process}
\end{equation}
This shows that $T^{x \rightarrow y}$ alone cannot capture any quantum interference effects encoded in the coherence terms $\rchi_{pq,rs}$ ($pq \neq rs$). Thus, it fails to characterize any coherence generation or preservation within $\{E_{pq}\}$.

To fully access quantum interference, a complementary stochastic matrix sensitive to coherence terms of \(\rchi\) is necessary~\bluecite{comple,complementary}. One choice of complementary basis [see Fig.~\hyperlink{fig:settings}{1(b)}] is
\begin{equation}
    \ket{m_u} = \frac{1}{\sqrt{d}} \sum_{k=0}^{d-1} \omega^{-km}\ket{k_x},
    \label{eq:complementary1}
\end{equation}
with $\omega = e^{2\pi i/d}$. The expected output states also form a complementary set:
\begin{equation}
    \ket{n_v} = \frac{1}{\sqrt{d}} \sum_{k=0}^{d-1} \omega^{-kn}\ket{k_y}.
    \label{eq:complementary2}
\end{equation}
The transition probability from input $m$ to outcome $n$ for this complementary experimental setting \((u,v)\) is given by:
\begin{align}
\nonumber
    T^{u \rightarrow v}_{mn} &= \bra{n_v}\mathcal{E}(\ketbra{m_u}{m_u})\ket{n_v}\\
    &=P(n_v| m_u).
    \label{eq:Tuv}
\end{align}
Substituting Eq.~\rref{eq:qpt} into Eq.~\rref{eq:Tuv}, we obtain
\begin{equation}
    T^{u \rightarrow v}_{mn} = \frac{1}{d^2} \sum_{p,q,r,s=0}^{d-1} \omega^{-(p - r)m + (q - s)n}\rchi_{pq,rs}.
    \label{eq:Tuv_chi}
\end{equation}
This result demonstrates that $T^{u \rightarrow v}$ constrains off-diagonal elements of $\rchi$. Together with $T^{x \rightarrow y}$, they capture critical information about incoherent and coherent processes in \(\mathcal{E}\).

While the preceding discussion focuses on the case where \(S\) is a unitary, AOQPT also applies to partial isometries. A partial isometry \(S\) acts unitarily from its initial space to its final space. The initial space is the orthogonal complement of its kernel, and the final space is its range. Importantly, a partial isometry \(S\) can be decomposed into a unitary evolution followed by an orthogonal projection~\bluecite{iso1, iso2}, as in quantum measurements with postselection. For \(k\)-dimensional initial and final spaces with \(k<d\), \(S\) preserves inner products and exhibits coherent evolution within these subspaces, effectively discarding components of quantum states outside them. Accordingly, \(T^{x \rightarrow y}_{mn}\) in Eq.~\rref{eq:Txy} is reconstructed by preparing input states \(\ket{m_x}\) spanning the initial space and measuring the output states with projectors \(\ketbra{n_y}\), where \(\ket{n_y}\) span the final space. Similarly, \(T^{u \rightarrow v}_{mn}\) in Eq.~\rref{eq:Tuv} is obtained using complementary bases of Eqs.~\rref{eq:complementary1} and~\rref{eq:complementary2}, with the normalization factor modified to \(1/\sqrt{k}\). This extension makes AOQPT suitable for analyzing relevant trace-nonincreasing operations, such as entanglement swapping~\bluecite{ent swap} and photonic fusion~\bluecite{fusion1}, the latter of which will be discussed in the experimental validation.

\begin{figure*}[t!]
    \centering
    \includegraphics{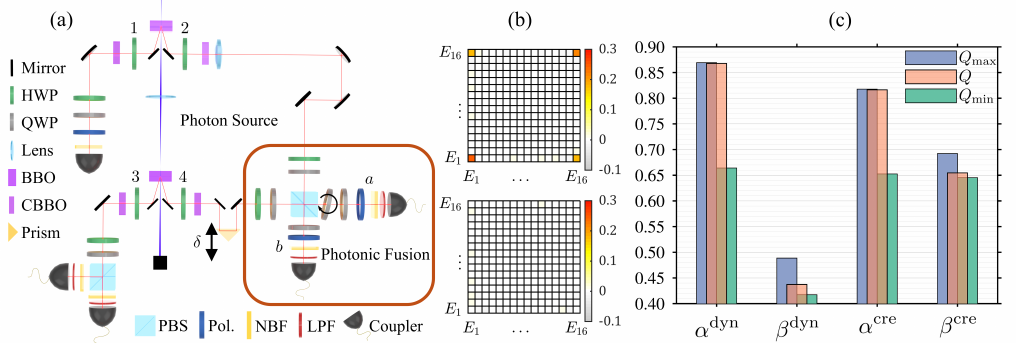}
    \caption{Experimental validation of AOQPT on photonic fusion. (a) Experimental setup for photonic fusion. Two heralded single-photon sources based on type-II spontaneous parametric down-conversion (SPDC) in \(\beta\)-barium-borate (BBO) crystals are pumped by a ultraviolet pulsed laser beam (center wavelength \(390\) nm, pulse duration \(144\) fs, and repetition rate 76 MHz). To compensate for spatial walk-off effects, each SPDC source incorporates an additional pair of compensation BBO (CBBO) crystals. Photon pairs are generated in spatial modes \((1,2)\) and \((3,4)\). Detection of vertically polarized photons in modes \((1,3)\)---filtered by a polarizer (Pol.) and a polarizing beam splitter (PBS)---heralds the presence of horizontally polarized single photons in modes \((2,4)\), respectively. To render photons in modes \((2,4)\) indistinguishable, the optical path of the photon in mode \(4\) is fine-tuned by adjusting the delay \(\delta\) via precise movement of the prism using a nanopositioning stage. To prepare different input states for AOQPT, we use a combination of a half-wave plate (HWP) and a quarter-wave plate (QWP). Projective measurements are realized by a QWP and a polarizer. The phase between photons in modes \((a,b)\) is adjusted by tilting a QWP in mode \(a\). Photons are filtered by a narrow-band filter (NBF) and a long-pass filter (LPF), collected by fiber couplers, and detected by single-photon avalanche diodes. (b) Experimentally reconstructed subnormalized process matrices $\rchi_\text{fusion}$ for the photonic fusion process, with a trace corresponding to a success probability of 1/2. The top and bottom matrices show the real and imaginary parts, respectively. \(\rchi_\text{fusion}\) is represented in the operator basis \(\{E_j\}\) with \(E_0=\ketbra{00}\), \(E_1=\ketbra{01}{00}\), \(E_2=\ketbra{10}{00}\), \(\ldots\), and \(E_{15}=\ketbra{11}\). The measured process fidelity is \(0.8155\). (c) Experimental results of SQPT and AOQPT applied to resource quantification for \(\rchi_\text{fusion}\). The resource measures are shown for both quantum-mechanical dynamics (\(\alpha^{\text{dyn}}, \beta^{\text{dyn}}\)) and entanglement-creation capability (\(\alpha^{\text{cre}}, \beta^{\text{cre}}\)). \(Q\) are obtained through the standard QPC method; \(Q_\text{max}\) and \(Q_\text{min}\) are computed through AOQPT.}
    \label{fig:result}
\end{figure*}

By reconstructing the complementary stochastic matrices, \((T^{x \rightarrow y}, T^{u \rightarrow v})\) as illustrated in Fig.~\hyperlink{fig:settings}{1(b)}, we define a decision process matrix \(\rchi'\) following the conditions defined by \((T^{x \rightarrow y}, T^{u \rightarrow v})\). \(\rchi'\) is then substituted into Eq.~\rref{Q} to estimate dynamical property measure \(Q\), yielding the following optimization problem:
\begin{equation}
\begin{aligned}
\text{given}\quad & \{T^{a \rightarrow b} | (a,b) = (x,y), (u,v)\} \\
\min_{\rchi'}\quad & C_Q(\rchi') = Q_\text{min} \leq Q \leq Q_\text{max} = \max_{\rchi'} C_Q(\rchi') \\
\text{s.t.}\quad 
& \bra{n_b} \mathcal{E}'(\ketbra{m_a}) \ket{n_b} = T^{a \rightarrow b}_{mn} \quad \forall m,n,\quad
 \rchi' \succeq 0.
\end{aligned}
\label{Qmit}
\end{equation}
The first constraint ensures that the transition probabilities of \(\mathcal{E}'\), characterized by \(\rchi'\), reproduce those of \(\mathcal{E}\), and the positive semidefinite constraint guarantees that \(\rchi'\) is a completely positive map. With this structure, Eq.~\rref{Qmit} can be naturally formulated as a semidefinite programming (SDP) problem, which is widely used to evaluate various dynamical properties, including process fidelity~\bluecite{fidelity,Hofmann}, diamond norm~\bluecite{dia1,dia2}, composition~\bluecite{quantify,capability}, and robustness~\bluecite{rob1,rob2,quantify,capability}. This method bounds \(Q\) using only two experimental settings, independent of system size, while mitigating artifacts from state preparation and measurement in conventional tomography. As each matrix describes \(d^2\) transition probabilities, Eq.~\rref{Qmit} imposes about \(2d^2\) independent probability constraints. Hence, compared with full characterization requiring \(d^4\) parameters, AOQPT achieves a \(1/d^2\) asymptotic cost reduction.

AOQPT provides a resource-efficient framework for quantifying dynamical properties by computing both the lower and upper bounds \((Q_\text{min}, Q_\text{max})\). For a resource measure \(Q\) satisfying conditions such as faithfulness and monotonicity~\bluecite{correlation,capability,channelresource1}, \(Q_{\min} > 0\) guarantees that the process \(\mathcal{E}\) possesses the corresponding dynamical resource (e.g., entanglement generation), whereas \(Q_{\max} = 0\) certifies its absence, allowing \(\mathcal{E}\) to be discarded early. Solving Eq.~\rref{Qmit} further yields the worst- and optimal-case process matrices within the set of valid \(\rchi'\) consistent with the measured data: the worst-case process reveals the minimum resource performance achievable under the given statistics, while the optimal-case process identifies the maximum potential performance allowed by the same data. This formulation further enables a modular analysis of large-scale quantum devices and networks~\bluecite{mod1,mod2,mod3}, where each experimental process \(\rchi\) can be replaced by its corresponding worst- or optimal-case counterpart from the feasible set of \(\rchi'\), thereby approximating the overall dynamics of a device or network. Hence, AOQPT provides a scalable tool for assessing the potential of quantum dynamics in complex systems.

\textit{Experimental validation.}---We here demonstrate the quantification of dynamical resources in a photonic fusion process, as a representative partially isometric operation. The analysis is conducted within the quantum process capability (QPC) framework~\bluecite{capability, quantify}, wherein a process is deemed \textit{incapable} if it lacks the ability to generate or preserve the resource, and \textit{capable} if it possesses such ability. We focus on two resource measures: composition, \(Q=\alpha\), which quantifies the minimal amount of a capable process present within the experimental process, and robustness, \(Q=\beta\), which quantifies the minimal amount of noise that can be added to render the process incapable (see Sec.~II of SM~\bluecite{SM}). Two dynamical resources are considered. The first, $Q^{\mathrm{dyn}}$, quantifies the ability to exhibit quantum-mechanical dynamics. Incapable processes are described using a classical picture: the initial system is specified by classical objects with properties satisfying realism, followed by classical stochastic evolution described by transition probabilities from classical objects into quantum states (see Sec.~II A of SM~\bluecite{SM}). The second resource, $Q^{\mathrm{cre}}$, quantifies the ability to create entanglement, with incapable processes being those that preserve separability for all separable input states (see Sec.~II B of SM~\bluecite{SM}). For completeness, an additional experimental validation is presented in Sec.~III of SM~\bluecite{SM}, using the \texttt{ibmq\_belem} device to analyze a CNOT gate as a representative unitary operation.

Photonic fusion \bluecite{fusion1,fusion2,fusion3,fusion4,fusion5,senior} is a fundamental primitive in optical quantum networks, generating multipartite entanglement and Bell state measurements that are essential for entanglement swapping~\bluecite{ent swap}, teleportation~\bluecite{tele}, and both measurement-based~\bluecite{mbqc1,mbqc2,mbqc3,mbqc4} and fusion-based~\bluecite{fbqc1,fbqc2,fbqc3} quantum computing. Figure~\hyperlink{fig:result}{2(a)} illustrates an photonic fusion. The PBS, a linear optical element acting as a \textit{unitary map}, transmits horizontal \(|H\rangle\) and reflects vertical \(|V\rangle\) polarization states. Conventionally, \(| H\rangle\) and \(| V\rangle\) are encoded as \(| 0\rangle\) and \(| 1 \rangle\), respectively. Postselecting two-fold coincidences in modes $a$ and $b$ corresponds to an \textit{orthogonal projection} onto the \(|0_a0_b\rangle\) and \(|1_a1_b\rangle\) basis states, which represent two different yet indistinguishable alternatives in the two-fold coincidence detection. Hence, the ideal photonic fusion process is a partial isometry, described by the Kraus operator \(S_\text{fusion} = |0_a0_b\rangle\langle 0_40_2| + |1_a1_b\rangle\langle 1_21_4|\), which acts on an input state \(\rho\) as \(S_\text{fusion} \rho S_\text{fusion}^\dagger\).

In the standard QPC method, the experimental process of the photonic fusion operation \(\mathcal{E}_\text{fusion}\) must be fully characterized by reconstructing its process matrix \(\rchi_\text{fusion}\) via SQPT [see Fig.~\hyperlink{fig:settings}{1(a)}]~\bluecite{procedure}. Once reconstructed, \(\rchi_\text{fusion}\) [Fig.~\hyperlink{fig:result}{2(b)}] is substituted into Eq.~\rref{Q} to compute the resource measures \(Q^\text{dyn}\) and \(Q^\text{cre}\). Figure~\hyperlink{fig:result}{2(c)} shows that the photonic fusion operation exhibits both quantum-mechanical dynamics and entanglement-creation capability. However, the standard QPC method requires \(81\) experimental settings, demonstrating its high experimental and computational cost.

In contrast, Fig.~\hyperlink{fig:settings}{1(b)} shows AOQPT requires only two complementary sets of stochastic matrices $(T^{x \rightarrow y}, T^{u \rightarrow v})$ to compute the capability bounds \((Q_\text{min},Q_\text{max})\) in Eq.~\rref{Qmit}. To implement AOQPT, the input states \(\{\ket{m_x}\}\) and the expected output states \(\{\ket{n_y}\}\) are chosen from the set \(\{\ket{00}, \ket{11}\}\), as it spans both the initial and the final spaces of \(S_\text{fusion}\). The corresponding stochastic matrix \(T^{x \rightarrow y}\) is then obtained with the experimental setting \((x,y)\). For the complementary experimental setting  \((u,v)\), the input states \(\{\ket{m_u}\}\) and outcomes \(\{\ket{n_v}\}\) are selected from the Bell basis \(\{\ket{\Phi^\pm}\}\), where \(\ket{\Phi^\pm} = (\ket{00} \pm \ket{11})/\sqrt{2}\). Theoretically, estimating \(Q\) via AOQPT can thus be estimated using only two experimental settings. However, to avoid the experimental complexity of directly preparing and measuring Bell states, \(\ketbra{\Phi^\pm}\) are decomposed as \((\mathbb{1} \pm XX \mp YY + ZZ)/4\), which requires 9 additional distinct Pauli experimental settings for constructing \(T^{u \rightarrow v}\). Consequently, AOQPT requires 10 experimental settings for photonic fusion, achieving an approximately \(88\%\) reduction in experimental cost.

After collecting the required transition probabilities, maximum likelihood estimation is applied to ensure physical validity and to construct $(T^{x \rightarrow y}, T^{u \rightarrow v})$. Substituting $(T^{x \rightarrow y}, T^{u \rightarrow v})$ into Eq.~\rref{Qmit} yields the lower and upper bounds \((Q_\text{min},Q_\text{max})\) of resource measures \(Q^\text{dyn}\) and \(Q^\text{cre}\). The experimental results illustrated in Fig.~\hyperlink{fig:result}{2(c)} shows that AOQPT provides tight bounds on resource measures for photonic fusion, closely matching the actual values---particularly for \(\alpha^\text{dyn}_\text{max}\), \(\alpha^\text{cre}_\text{max}\), and \(\beta^\text{cre}_\text{min}\). While \(\alpha^\text{dyn}_\text{min}\), \(\beta^\text{dyn}_\text{max}\), and \(\alpha^\text{cre}_\text{min}\) are conservative, the true values lie well within AOQPT bounds. These results demonstrate that AOQPT offers accurate and efficient certification of quantum-mechanical dynamics and entanglement-creation capability.

\textit{AOQPT applications.}---For most quantum information tasks, Eq.\rref{Qmit} can be formulated as a SDP. Although the computational cost scales exponentially with system size, AOQPT remains tractable for systems of up to \(5\) qubits. These SDPs may involve millions of variables and constraints, yet are solvable with current computational tools\bluecite{modern1, modern2, modern3}. This makes AOQPT a viable method for diverse quantum technologies.

One application is in chip-scale quantum processors~\bluecite{chip1,chip2}, where AOQPT can estimate process fidelity for universal logic gates using a fixed number of experimental settings~\bluecite{Hofmann}. It can also bound the diamond norm~\bluecite{dia1,dia2}, which determines the success probability in single-shot channel discrimination~\bluecite{dia}, and thus allows this probability to be estimated efficiently. Beyond these, AOQPT can assess the quality of multipartite entangled states generation, as demonstrated in the experimental validation. Hence, it can estimate a process’s capability to create and manipulate cluster states~\bluecite{cluster1,cluster2}, which are central to both measurement-based~\bluecite{mbqc1,mbqc2,mbqc3,mbqc4} and fusion-based~\bluecite{fbqc1,fbqc2,fbqc3} quantum computing. As a system-size-independent tomographic method, AOQPT also provides a benchmarking tool for entanglement-based quantum networks with modular architectures~\bluecite{mod1,mod2,mod3}, supporting scalable quantum communication tasks like teleportation~\bluecite{tele} and distributed quantum computing~\bluecite{net3}. Requiring only two experimental settings, AOQPT not only establishes lower and upper bounds on dynamical properties but also characterizes the underlying processes across entire modular networks. Its system-size independence alleviates artifacts from imperfect state preparation and measurement, thereby benchmarking the intrinsic dynamics of quantum networks.

To extend AOQPT to completely unknown quantum processes, one can combine it with classical shadow tomography~\bluecite{classicalshadow}, which efficiently predicts properties of quantum systems such as expectation values, fidelities, and entanglement measures. In this way, input states are prepared from mutually unbiased bases, and randomized measurements or single-setting quantum state tomography~\bluecite{singleQST} are applied to the outputs to reconstruct classical shadows. When integrated with machine learning models, this hybrid AOQPT–shadow framework enables scalable prediction of dynamical properties, making it possible to characterize quantum dynamics far beyond the reach of conventional tomography.

\textit{Conclusions.}---We have proposed approximate optimization quantum process tomography (AOQPT) for estimating dynamical properties by computing their bounds and reconstructing the worst- and optimal-case processes for further analysis on quantum devices and networks. Given a target operation, this method enables system-size-independent quantification of dynamical properties by extracting essential information from two complementary experimental settings. Experimental validation on the photonic fusion operation and the CNOT gate demonstrates the reliability of AOQPT in resource estimation. Its applications include process fidelity estimation, diamond distance bounding, and the quantification of dynamical resources. These findings motivate several open questions. First, how can AOQPT be extended to accommodate larger quantum systems? Second, how can the method be applied to quantify dynamical properties when state preparation and measurement are error-prone or even untrusted? Third, how can AOQPT be generalized to a broader class of quantum operations beyond unitary and partially isometric operations?


\begin{thebibliography}{99}
\bibitem{ancilla} In ancilla-assisted process tomography~\bluecite{AAQPT}, the number of required settings remains \(3^{2N}\): \(d^2\) separable inputs are replaced by a single maximally entangled state from a \(d^2\)-dimensional Hilbert space, and quantum state tomography is performed on the corresponding output state.
\bibitem{orthogonality} Two operators \(A\) and \(B\) are said to be orthogonal if the Hilbert–Schmidt inner product \(\text{Tr}(A^\dagger B)=0\).
\bibitem{complementary} Two sets of quantum states are said to be complementary if the bases \(\{\ket{m_x}\}_{m=0}^{d-1}\) and \(\{\ket{n_u}\}_{n=0}^{d-1}\) are mutually unbiased: \(\abs{\bra{m_x}\ket{n_u}}^2=1/d\quad\forall m,n\).
\end{thebibliography}
\end{document}


\title{Supplemental Material for Quantification of Quantum Dynamical Properties with Two Experimental Settings}

\author{Tzu-Liang Hsu}
\author{Kuan-Jou Wang}
\author{Chun-Hao Chang}
\author{Sheng-Yan Sun}
\author{Shih-Husan Chen}
\author{Ching-Jui Huang}
\author{Che-Ming Li}
\email{cmli@mail.ncku.edu.tw}
\affiliation{Department of Engineering Science, National Cheng Kung University, Tainan 70101, Taiwan}

\affiliation{Center for Quantum Frontiers of Research and Technology, National Cheng Kung University, Tainan 70101, Taiwan}

\date{\today}
\begin{abstract}
In this Supplemental Material, we present quantum process tomography, with a focus on the procedure for reconstructing a process matrix. We also review quantum process capability theory, and derives its primal and dual semidefinite programming formulations for quantifying the ability to exhibit quantum-mechanical dynamics and generate entanglement. Furthermore, we demonstrate the method through an additional experimental validation of approximate optimization quantum process tomography on a controlled-NOT gate implemented on the 5-qubit \texttt{ibmq\_belem} device.
\end{abstract}

\maketitle


\section{Quantum Process Tomography}

In this section, we detail the mathematical framework and experimental procedures required to perform standard quantum process tomography (SQPT), with a particular focus on deriving the total number of experimental settings required to evaluate dynamical property measures as referenced in Eq.~(1) of the main text. The experimental process matrices of the photonic fusion operation [Fig.~2(b) of the main text] and the controlled-NOT (CNOT) gate (Fig.~\ref{fig:chi_plots}) are reconstructed according to the following procedures.

A general quantum operation \(\mathcal{E}\), acting on a \(d\)-dimensional Hilbert space \(\mathcal{H}\) and applied to a quantum state \(\rho\), can be described using the operator-sum (Kraus) representation~\bluecite{QCQI, chi1, chi2}:
\begin{equation}
    \mathcal{E}(\rho) = \sum_{k=0}^{d^2 -1} A_k \rho A_k^\dagger,
    \label{eq:S1}
\end{equation}
where \(\{A_k\}\) is a set of Kraus operators. These operators capture the physical action of the process and must satisfy the completeness condition:
\begin{equation}
    \sum_k A_k^\dagger A_k \preceq \mathbb{1}.
    \label{eq:S2}
\end{equation}
Equality holds if and only if \(\mathcal{E}\) is trace-preserving (TP), ensuring that the output state remains properly normalized. To reconstruct the process \(\mathcal{E}\) from measurable data, each Kraus operator \(A_k\) can be expanded in a fixed operator basis \(\{E_m\}_{m=0}^{d^2 - 1}\), which spans the space of linear operators on the Hilbert space:
\begin{equation}
    A_k = \sum_{m=0}^{d^2 - 1} a_{km} E_m,
    \label{eq:S3}
\end{equation}
where \(a_{km}\) are complex coefficients. Substituting Eq.~\eqref{eq:S3} into Eq.~\eqref{eq:S1} yields a \(\rchi\)-matrix representation:
\begin{equation}
    \mathcal{E}(\rho) = \sum_{m,n=0}^{d^2 - 1} \rchi_{mn} E_m \rho E_n^\dagger,
    \label{eq:S4}
\end{equation}
where the matrix elements are defined as \(\rchi_{mn} = \sum_k a_{km} a_{kn}^*\). \(\rchi\) thus is Hermitian and positive semidefinite by construction. The TP condition in Eq.~\eqref{eq:S2} imposes a linear constraint on \(\rchi\):
\begin{equation}
    \sum_{m,n=0}^{d^2 - 1} \rchi_{mn} E_m^\dagger E_n \preceq \mathbb{1}.
    \label{eq:S5}
\end{equation}
Therefore, fully characterizing a quantum operation \(\mathcal{E}\) requires determining the \(d^4\) real parameters that describe \(\rchi\). If the map is TP, Eq.~\eqref{eq:S5} imposes \(d^2\) real constraints, reducing the number of independent parameters to \(d^4 - d^2\).

SQPT aims to characterize an unknown quantum operation \(\mathcal{E}\) by preparing a complete basis of input states and performing full quantum state tomography (QST) on the corresponding output states. To achieve this, it requires preparing a set of \(d^2\) linearly independent states \(\{\rho_j\}_{j=0}^{d^2-1}\), which span the space of linear operators on a \(d\)-dimensional Hilbert space \(\mathcal{H}\). In practice, one can simply chose \(\rho_j = \ketbra{m}{n}\), with \(m,n\in\{0,\ldots,d-1\}\), and these output basis states \(\mathcal{E}(\rho_j)\) can then be produced experimentally by preparing the input states \(\ket{m}\), \(\ket{n}\), \(\ket{+}=(\ket{m}+\ket{n})/\sqrt{2}\), and \(\ket{+i}=(\ket{m}+i\ket{n})/\sqrt{2}\) and following the linear combinations:
\begin{align}
    \nonumber \mathcal{E}(\rho_j)&=
    \mathcal{E}(\ketbra{m}{n})\\
    \nonumber&=\mathcal{E}(\ketbra{+})+i\mathcal{E}(\ketbra{+i})-\frac{1+i}{2}\mathcal{E}(\ketbra{m})-\frac{1+i}{2}\mathcal{E}(\ketbra{n}).
\end{align}
Here the output states are \(\mathcal{E}(\ketbra{m})\), \(\mathcal{E}(\ketbra{n})\), \(\mathcal{E}(\ketbra{+})\), and \(\mathcal{E}(\ketbra{+i})\) are reconstructed via QST. Subsequently, the output basis states can then be expanded in the same operator basis \(\{\rho_k\}\) as
\begin{equation}
    \mathcal{E}(\rho_j) = \sum_{k=0}^{d^2 - 1} \lambda_{jk} \rho_k.
    \label{eq:S6}
\end{equation}
The coefficients \(\lambda_{jk}\) can be directly obtained by
\begin{equation}
    \lambda_{jk} = \Tr\left[ \mathcal{E}(\rho_j) \rho_k \right],
    \label{eq:S7}
\end{equation}
where \(\lambda_{jk}\) quantifies the weight of the basis operator \(\rho_k\). To connect this expression with the \(\rchi\)-matrix representation in Eq.~\eqref{eq:S4}, we express \(E_m \rho_j E_n^\dagger\) as a linear combination of the basis states \(\rho_k\),
\begin{equation}
    E_m \rho_j E_n^\dagger = \sum_{k=0}^{d^2 - 1} B_{mn,jk} \rho_k.
    \label{eq:S8}
\end{equation}
Substituting Eqs.~\eqref{eq:S8} into~\eqref{eq:S4}, the output becomes
\begin{equation}
    \mathcal{E}(\rho_j) = \sum_{m,n=0}^{d^2 - 1} \rchi_{mn} \sum_{k=0}^{d^2 - 1} B_{mn,jk} \rho_k 
    = \sum_{k=0}^{d^2 - 1} \left( \sum_{m,n=0}^{d^2 - 1} \rchi_{mn} B_{mn,jk} \right) \rho_k,
    \label{eq:S9}
\end{equation}
which implies the relationship
\begin{equation}
    \lambda_{jk} = \sum_{m,n=0}^{d^2 - 1} B_{mn,jk} \rchi_{mn}.
    \label{eq:S10}
\end{equation}
To proceed, the coefficient matrix \(\lambda\) and the process matrix \(\rchi\) can each be reshaped into \(d^2\)-dimensional vectors, while the tensor \(B\) can be regarded as a \(d^2 \times d^2\) matrix, with columns indexed by the pair \((mn)\) and rows by \((jk)\). This reformulation yields a linear system of the form \(\lambda = B \rchi\). If \(K\) denotes the generalized inverse of \(B\), satisfying
\[
\sum_{j,k} K_{pq,jk} B_{mn,jk} = \delta_{pm} \delta_{qn},
\]
then the elements of the process matrix \(\rchi\) can be recovered as
\begin{equation}
    \rchi_{mn} = \sum_{j,k=0}^{d^2 - 1} K_{mn,jk} \lambda_{jk}.
    \label{eq:S11}
\end{equation}

In the case of \(N\)-qubit systems, with \(d=2^N\), QPT is exclusively performed using Pauli settings. The prepared input states \(\rho_\text{in}=\ketbra{m},\ketbra{n},\ketbra{+},\ketbra{+i}\) are the eigenstates of \(N\)-qubit Pauli operators \(\{P_k\}_{k=0}^{3^N-1}\) using \(3^N\) settings. For each input state \(\rho_\text{in}\), the output state \(\mathcal{E}(\rho_\text{in})\) is reconstructed using Pauli settings:
\begin{equation}
    \mathcal{E}(\rho_\text{in})=\sum_{k=0}^{3^N-1}\frac{\Tr\bigl[P_k\mathcal{E}(\rho_\text{in})\bigr]P_k}{2^N}.
\end{equation}
Consequently, the total number of input-output experimental settings required for QPT is \(3^{2N}\), which scales exponentially with the system size and results in a severe experimental overhead.

\section{Quantum Process Capability}

In this section, we focus on two resource measures: composition \(\alpha\) and robustness \(\beta\) introduced in the quantum process capability (QPC) theory, as representative dynamical properties [Eqs.~(1) and~(10) of the main text]. Specifically, we consider two dynamical resources: the ability to exhibit quantum-mechanical dynamics and generate entanglement of quantum states, as detailed in Secs.~\ref{sec:dyn} and~\ref{sec:cre}, respectively.

To evaluate the resource measures defined in QPC theory via \textit{approximate optimization quantum process tomography} (AOQPT), we begin with the optimization problem given in Eq.~(10) of the main text: 
\begin{equation}
\begin{aligned}
\text{given}\quad & \{T^{a \rightarrow b} | (a,b) = (x,y), (u,v)\} \\
\min_{\rchi'}\quad & C_Q(\rchi') = Q_\text{min} \leq Q \leq Q_\text{max} = \max_{\rchi'} C_Q(\rchi') \\
\text{s.t.}\quad 
& \bra{n_b} \mathcal{E}'(\ketbra{m_a}) \ket{n_b} = T^{a \rightarrow b}_{mn} \quad \forall m,n,\quad
 \rchi' \succeq 0.
\end{aligned}
\label{Qmit}
\end{equation}
Here, \(C_Q \in \{C_\alpha, C_\beta\}\) represents a capability function quantifying a specific resource measure. The QPC framework adopts a deductive approach, comparing the experimental process \(\mathcal{E}\) with incapable processes \(\mathcal{E}_\text{I}\) that lack the ability to generate or preserve the static resource of interest~\bluecite{quantify,capability}.


The composition \(\alpha\) is defined to be the minimal amount of the capable process \(\mathcal{E}_\text{Q}\) present in \(\mathcal{E}\). Formally, the quantification problem in Eq.~\eqref{Qmit} for \(\alpha\) is given by \(\alpha=C_\alpha(\rchi)\):
\begin{equation}
\begin{aligned}
\alpha = \min_{c,\rchi_\text{I}, \rchi_\text{Q}}\quad&c \\
\text{s.t.}\quad
& \rchi = c\rchi_\text{Q}+(1-c)\rchi_\text{I}, \\
& 0\leq c\leq 1, \quad \rchi_\text{I}\in \mathcal{D}_{\rchi_\text{I}},\\
& \rchi_\text{Q} \succeq 0,\quad \rchi_\text{I} \succeq 0,
\end{aligned}
\label{primal_alpha}
\end{equation}
where \(c\) is the composition ratio, and \(\rchi_\text{Q}\) and \(\rchi_\text{I}\) characterize capable process \(\mathcal{E}_\text{Q}\) and incapable process \(\mathcal{E}_\text{I}\), respectively. The set \(  \mathcal{D}_{\rchi_\text{I}} \) consists of all \( \rchi_\text{I} \) satisfying the constraints that specify the properties of \( \rchi_\text{I} \)~\bluecite{capability}. The last two constraints ensure both \(\mathcal{E}_\text{Q}\) and \(\mathcal{E}_\text{I}\) remain CP maps. As for the robustness \(\beta\), it is defined to be the minimal amount of noise process required to make \(\rchi\) incapable. More precisely, the quantification problem in Eq.~\eqref{Qmit} for \(\beta\) is given by \(\beta = C_\beta(\rchi)\):
\begin{equation}
\begin{aligned}
\beta = \min_{r,\rchi_\text{I}, \rchi_\text{N}}\quad&r\\ 
\text{s.t.}\quad 
& \frac{\rchi+r\rchi_\text{N}}{1+r}=\rchi_\text{I}, \\
& r\geq 0,\quad \rchi_\text{I}\in  \mathcal{D}_{\rchi_\text{I}},\\
& \rchi_\text{N} \succeq 0, \quad \rchi_\text{I} \succeq 0,
\end{aligned}
\label{primal_beta}
\end{equation}
where \(r\) is the noise strength. The set \( \mathcal{D}_{\rchi_\text{I}}\) is the same as that in Eq.~\eqref{primal_alpha}. The last two constraints ensure both the noise process characterized by \(\mathcal{E}_\text{N}\) and \(\mathcal{E}_\text{I}\) are CP maps. The set \( \mathcal{D}_{\rchi_\text{I}}\) shown in Eqs.~\eqref{primal_alpha} and \eqref{primal_beta} can impose various conditions on the incapable process \(\rchi_\text{I}\). By specifying different sets of constraints \(\mathcal{D}\), we can address different problems of resource quantification via AOQPT. 

Once the set of constraints \( \mathcal{D}_{\rchi_\text{I}}\) defining the incapable process \(\mathcal{E}_\text{I}\) has been identified, the lower bounds of the QPC measures shown in Eqs.~\eqref{primal_alpha} and \eqref{primal_beta} can be evaluated numerically via semi-definite programming (SDP)~\bluecite{sdp}. As an SDP solver, such as MOSEK~\bluecite{MOSEK}, produces valid decision matrices \(\rchi'\) that inherently encode key information from the constraints in Eq.~\eqref{Qmit}, the lower bounds on the composition and robustness, denoted by \(\alpha_\text{min}\) and \(\beta_\text{min}\), respectively, can be computed directly by substituting Eqs.~\eqref{primal_alpha} and \eqref{primal_beta} into Eq.~\eqref{Qmit}. However, this approach does not apply to the upper bounds, which involve max-min optimization. Instead, they can be determined by deriving their dual formulations~\bluecite{quantumsteering, lecturenote, duality}. By introducing Lagrange variables for each constraint, under strong duality, the dual SDP for Eq.~\eqref{primal_alpha} is given by
\begin{equation}
\begin{aligned}
\alpha = \max_{F}\quad&1-\Tr\bigl(F\rchi\bigr), \\
\text{s.t.}\quad
& F \in \mathcal{D}_F,\quad
 F \succeq 0,
\end{aligned}
\label{dual_alpha}
\end{equation}
where \(F\) is a dual variable, and the set \(\mathcal{D}_F\) contains all \(F\) satisfying the necessary dual conditions. Similarly, the dual SDP for Eq.~\eqref{primal_beta} is
\begin{equation}
\begin{aligned}
\beta = \max_{G}\quad &\Tr\bigl(G\rchi\bigr) -1\\
\text{s.t.}\quad
& G \in \mathcal{D}_G,\quad
 G \succeq 0,
\end{aligned}
\label{dual_beta}
\end{equation}
where \(G\) is a dual variable subject to the necessary dual constraints in \(\mathcal{D}_G\). 

In conjunction with Eq.~\eqref{Qmit}, the dual objective function in Eq.~\eqref{dual_alpha} now is bilinear in the dual variable \(F\) and the process matrix \(\rchi'\), respectively. The constraints are decoupled into those involving \(F\) and those involving \(\rchi'\). Hence, this structure suggests that the upper bounds on the composition, denoted by \(\alpha_\text{max}\), respectively, can be calculated heuristically by applying a see-saw optimization algorithm: alternatively hold either \(F\) or \(\rchi'\) fixed. Starting from a feasible \(F^{(0)}\), the solver first generates the \(\rchi'^{(0)}\) maximizing the dual objective function with constraints in Eq.~\eqref{Qmit}. Next, the measures are updated by searching for the best feasible \(F^{(1)}\) satisfying constraints in Eq.~\eqref{dual_alpha}. Finally, this procedure is iterated to generate the sequence \(\{F^{(i)},\rchi'^{(i)}\}\) until convergence or numerical precision is reached. In the same way, the upper bound on robustness, \(\beta_\text{max}\), can be computed by applying the see-saw algorithm, with Eq.~\eqref{dual_alpha} replaced by Eq.~\eqref{dual_beta}.

In the next subsections, we study the two target dynamical resources---ability to exhibit quantum-mechanical dynamics and to generate entanglement---by explicitly defining the constraints of their incapable processes and present their corresponding primal and dual SDP formulations following Eqs.~\eqref{primal_alpha} to \eqref{dual_beta}.

\subsection{Quantum-Mechanical Dynamics}\label{sec:dyn}
When quantifying quantum-mechanical dynamics of \(\rchi\)~\bluecite{quantify}, an incapable process \(\mathcal{E}_\text{I}\) can be explained using a classical picture: the initial system is prescribed by classical objects \(\mathbf{v}_\mu\) with properties \(a_x\) satisfying realism. The system then undergoes classical stochastic evolution described by transition probabilities \(\Omega_{\xi\mu}\) from \(\mathbf{v}_\xi\) into quantum states \(\sigma_\mu\). Accordingly, the set of constraints in Eqs.~\eqref{primal_alpha} and \eqref{primal_beta} for \(\mathcal{E}_\text{I}\) can be defined to be 
\begin{equation}
    \mathcal{D}_{\rchi_\text{I}}=\Bigl\{\rchi_\text{I}| \mathcal{E}_\text{I}(\sigma_{a|x})=\sum_{\xi,\mu} p(\mathbf{v}_\xi |a_x)\Omega_{\xi\mu}\sigma_\mu \quad \forall a,x\Bigr\},
    \label{c1}
\end{equation}
where the conditional states \(\sigma_{a|x}\) are eigenstates of \(\hat{a}_x\) with eigenvalues \(a\), and the set \(\{\sigma_{a|x}\}\) is tomographically complete for QPT. The classical stochastic dynamics can always be decomposed as a convex combination of deterministic transitions from a classical object \(\mathbf{v}_\xi\) to a quantum state \(\rho_{\mu'}\):
\begin{equation}
    \Omega_{\xi\mu}=\sum_{\mu'}p(\mu'|\mu)\delta_{\xi\mu'},
    \label{det}
\end{equation}
where \(\delta_{\xi\mu'}\) is the deterministic transition probability, and \(p(\mu'|\mu)\) is the weight of the deterministic transition probability labeled by \(\mu'\). Substituting Eq.~\eqref{det} into Eq.~\eqref{c1} and defining \(\rho_{\mu'}=\sum_\mu p(\mu'|\mu)\sigma_\mu\), the output state of \(\mathcal{E}_\text{I}\) can be expressed as
\begin{equation}
    \mathcal{E}_\text{I}(\sigma_{a|x})=\sum_{\mu'}p(\mathbf{v}_{\mu'} |a_x)\rho_{\mu'}.
    \label{fin}
\end{equation}
Assuming a uniform distribution over \(p(a_x)\), Bayes' rule allows the conditional probability to be rewritten as
\(
p(\mathbf{v}_{\mu'} | a_x) = d \cdot p(\mathbf{v}_{\mu'}) p(a_x | \mathbf{v}_{\mu'}).
\)
Substituting this into Eq.~\eqref{fin} and relabeling \(\mu'\rightarrow\mu\), the output of the incapable process \(\mathcal{E}_\text{I}\) therefore becomes
\begin{equation}
    \mathcal{E}_\text{I}(\sigma_{a|x}) = d \sum_{\mu} p(\mathbf{v}_\mu) p(a_x | \mathbf{v}_\mu) \rho_\mu \quad \forall a,x.
    \label{eq:S13}
\end{equation}

By inserting Eq.~\eqref{eq:S13} into Eq.~\eqref{primal_alpha}, the quantum-mechanical dynamics composition \(\alpha^\text{dyn}\) can be obtained by solving the following optimization problem:
\begin{equation}
\begin{aligned}
\alpha^\text{dyn} = \min_{c^\text{dyn}, \rchi_\text{Q}, \rchi_\text{I},\rho_\mu} \quad & c^\text{dyn} \\
\text{s.t.} \quad 
& \rchi = c^\text{dyn} \rchi_\text{Q} + (1 - c^\text{dyn}) \rchi_\text{I}, \\
& 0 \leq c^\text{dyn} \leq 1, \quad
 \mathcal{E}_\text{I}(\sigma_{a|x}) = d \sum_{\mu} p(\mathbf{v}_\mu) 
   p(a_x | \mathbf{v}_\mu) \rho_\mu\quad \forall a,x, \\
& \rchi_\text{Q} \succeq 0,\quad \rchi_\text{I} \succeq 0.
\end{aligned}
\label{eq:S14}
\end{equation}

To express this as an SDP, we define an unnormalized incapable process \(\Tilde{\mathcal{E}}_\text{I} = (1 - c^\text{dyn}) \mathcal{E}_\text{I}\), characterized by the unnormalized process matrix \(\Tilde{\rchi}_\text{I} = (1 - c^\text{dyn}) \rchi_\text{I}\) and unnormalized quantum states \(\Tilde{\rho}_\mu = (1 - c^\text{dyn}) p(\mathbf{v}_\mu) \rho_\mu\). The resulting SDP formulation becomes:
\begin{equation}
\begin{aligned}
\alpha^\text{dyn} = \min_{\Tilde{\rchi}_\text{I}, \Tilde{\rho}_\mu} \quad& 1 - \Tr\left( \Tilde{\rchi}_\text{I} \right) \\
\text{s.t.} \quad
& \Tilde{\mathcal{E}}_\text{I}(\sigma_{a|x}) = d \sum_{\mu} p(a_x | \mathbf{v}_\mu) \Tilde{\rho}_\mu \quad \forall a,x, \\
& \rchi - \Tilde{\rchi}_\text{I} \succeq 0, \quad \Tilde{\rchi}_\text{I} \succeq 0, \quad \Tilde{\rho}_\mu \succeq 0 \quad \forall \mu.
\end{aligned}
\label{eq:S15}
\end{equation}

The SDP formulation in Eq.~\eqref{eq:S15} involves the action of the unnormalized incapable process \(\Tilde{\mathcal{E}}_\text{I}\) on the input \(\sigma_{a|x}\), which can be conveniently expressed in terms of \(\rchi\). By the Choi--Jamiolkowski isomorphism \bluecite{choi}, this action takes the form
\begin{equation}
    \Tilde{\mathcal{E}}_\text{I}(\sigma_{a|x}) = d \Tr_1\left[ \left( \sigma_{a|x}^{\text{T}} \otimes \mathbb{1} \right) \Tilde{\rchi}_\text{I} \right],
    \label{eq:S16}
\end{equation}
where \(\Tilde{\rchi}_\text{I}\) is defined in the computational basis, \(\text{T}\) denotes matrix transposition, and \(\Tr_1\) indicates the partial trace over the first subsystem. To derive the dual of the SDP in Eq.~\eqref{eq:S15}, we introduce dual variables associated with each constraint \bluecite{lecturenote, quantumsteering}. The corresponding Lagrangian function is
\begin{align}
\mathcal{L} 
&= 1 - \Tr\left( \Tilde{\rchi}_\text{I} \right) 
- \sum_{a,x} \Tr\left\{ K_{a|x} \left[ \Tr_1\left( \left( \sigma_{a|x}^{\text{T}} \otimes \mathbb{1} \right) \Tilde{\rchi}_\text{I} \right) - \sum_\mu p(a_x | \mathbf{v}_\mu) \Tilde{\rho}_\mu \right] \right\} \notag \\
&\quad - \Tr\left[ F \left( \rchi - \Tilde{\rchi}_\text{I} \right) \right] 
- \Tr\left( M \Tilde{\rchi}_\text{I} \right)
- \sum_\mu \Tr\left( N_\mu \Tilde{\rho}_\mu \right) \notag \\
&= 1 - \Tr(F \rchi) 
+ \Tr\left[ \left( F - \sum_{a,x} \sigma_{a|x}^{\text{T}} \otimes K_{a|x} - \mathbb{1} - M \right) \Tilde{\rchi}_\text{I} \right] + \sum_\mu \Tr\left[ \left( \sum_{a,x} p(a_x | \mathbf{v}_\mu) K_{a|x} - N_\mu \right) \Tilde{\rho}_\mu \right],
\label{eq:S17}
\end{align}
where the dual variables satisfy the positivity constraints: \(F \succeq 0\), \(M \succeq 0\), \(N_\mu \succeq 0\), and \(K_{a|x} = K_{a|x}^\dagger\). The Lagrangian \(\mathcal{L}\) becomes independent of the primal variables \(\Tilde{\rchi}_\text{I}\) and \(\Tilde{\rho}_\mu\) when the following  conditions are imposed:
\begin{equation}
    \nonumber
    M = F - \sum_{a,x} \sigma_{a|x}^{\text{T}} \otimes K_{a|x} - \mathbb{1}, \quad    
    N_\mu = \sum_{a,x} p(a_x | \mathbf{v}_\mu) K_{a|x}.
\label{eq:S18}
\end{equation}
Substituting these expressions back into \(\mathcal{L}\) yields the dual SDP:
\begin{equation}
\begin{aligned}
\alpha^\text{dyn} = \max_{F, K_{a|x}} \quad& 1 - \Tr\left( F \rchi \right) \\
\text{s.t.} \quad
&  F - \sum_{a,x} \sigma_{a|x}^{\text{T}} \otimes K_{a|x} - \mathbb{1} \succeq 0,\\
& \sum_{a,x} p(a_x | \mathbf{v}_\mu) K_{a|x} \succeq 0 \quad \forall \mu, \quad
 F \succeq 0.
\end{aligned}
\label{eq:S19}
\end{equation}

For the quantum-mechanical dynamics robustness \(\beta^\text{dyn}\), inserting Eq.~\eqref{eq:S13} into Eq.~\eqref{primal_beta} from the main text yields the following optimization problem:
\begin{equation}
\begin{aligned}
\beta^\text{dyn} = \min_{r^\text{dyn}, \rchi_\text{N},\rchi_\text{I},\rho_\mu} \quad& r^\text{dyn} \\
\text{s.t.} \quad& 
\frac{\rchi + r^\text{dyn} \rchi_\text{N}}{1 + r^\text{dyn}} = \rchi_\text{I}, \\
& r^\text{dyn} \geq 0, \quad
 \mathcal{E}_\text{I}(\sigma_{a|x}) = d \sum_{\mu} p(\mathbf{v}_\mu) p(a_x | \mathbf{v}_\mu) \rho_\mu \quad \forall a,x, \\
& \rchi_\text{N} \succeq 0, \quad \rchi_\text{I} \succeq 0.
\end{aligned}
\label{eq:S20}
\end{equation}
Introducing unnormalized variables
\(
\Tilde{\rchi}_\text{I} = (1 + r^\text{dyn}) \rchi_\text{I}\) and \(\Tilde{\rho}_\mu = (1 + r^\text{dyn}) p(\mathbf{v}_\mu) \rho_\mu,
\)
the problem can be reformulated as the following SDP problem:
\begin{equation}
\begin{aligned}
\beta^\text{dyn} = \min_{\Tilde{\rchi}_\text{I}, \Tilde{\rho}_\mu} \quad& \Tr\left( \Tilde{\rchi}_\text{I} \right) - 1 \\
\text{s.t.} \quad& 
\Tilde{\mathcal{E}}_\text{I}(\sigma_{a|x}) = d \sum_{\mu} p(a_x | \mathbf{v}_\mu) \Tilde{\rho}_\mu \quad \forall a,x, \\
& \Tilde{\rchi}_\text{I} - \rchi \succeq 0, \quad \Tilde{\rchi}_\text{I} \succeq 0, \quad \Tilde{\rho}_\mu \succeq 0 \quad \forall \mu.
\end{aligned}
\label{eq:S21}
\end{equation}

When the output is computed via the Choi--Jamiolkowski representation as introduced in Eq.~\eqref{eq:S16}, the Lagrangian associated with this primal problem reads:
\begin{align}
\mathcal{L} &= \Tr\left( \Tilde{\rchi}_\text{I} \right) - 1 
- \sum_{a,x} \Tr\left\{ K_{a|x} \left[ \Tr_1\left( \left( \sigma_{a|x}^{\text{T}} \otimes \mathbb{1} \right) \Tilde{\rchi}_\text{I} \right) - \sum_\mu p(a_x | \mathbf{v}_\mu) \Tilde{\rho}_\mu \right] \right\} \notag \\
&\quad - \Tr\left[ G \left( \Tilde{\rchi}_\text{I} - \rchi \right) \right]
- \Tr\left( M \Tilde{\rchi}_\text{I} \right)
- \sum_\mu \Tr\left( N_\mu \Tilde{\rho}_\mu \right) \notag \\
&= \Tr(G \rchi) - 1 
+ \Tr\left[ \left( \mathbb{1} - G - \sum_{a,x} \sigma_{a|x}^{\text{T}} \otimes K_{a|x} - M \right) \Tilde{\rchi}_\text{I} \right] + \sum_\mu \Tr\left[ \left( \sum_{a,x} p(a_x | \mathbf{v}_\mu) K_{a|x} - N_\mu \right) \Tilde{\rho}_\mu \right],
\label{eq:S22}
\end{align}

To eliminate dependence on the primal variables \(\Tilde{\rchi}_\text{I}, \Tilde{\rho}_\mu\), we impose the following conditions:
\begin{equation}
\nonumber M = \mathbb{1} - G - \sum_{a,x} \sigma_{a|x}^{\text{T}} \otimes K_{a|x},\quad
N_\mu = \sum_{a,x} p(a_x | \mathbf{v}_\mu) K_{a|x}.
\label{eq:S23}
\end{equation}
Substituting these into \(\mathcal{L}\) yields the dual SDP:
\begin{equation}
\begin{aligned}
\beta^\text{dyn} = \max_{G, K_{a|x}} \quad& \Tr(G \rchi) - 1 \\
\text{s.t.} \quad& 
\mathbb{1} - G - \sum_{a,x} \sigma_{a|x}^{\text{T}} \otimes K_{a|x} \succeq 0,\\
&\sum_{a,x} p(a_x | \mathbf{v}_\mu) K_{a|x} \succeq 0 \quad \forall \mu,  \quad G \succeq 0.
\end{aligned}
\label{eq:S24}
\end{equation}

\subsection{Entanglement-Creation Capability}\label{sec:cre}
Another example is the entanglement-creation capability~\bluecite{capability}, where an incapable process \(\mathcal{E}_\text{I}\) preserves separability only: each output state remains separable for every separable input state. For a two-qubit system, the output states follow the Peres-Horodecki criterion~\bluecite{ppt1,ppt2}, so the set of constraints for such process acting on a separable states \(\rho_s\) is defined to be
\begin{equation}
    \mathcal{D}_{\rchi_\text{I}}=\Bigl\{\rchi_\text{I}|\mathcal{E}_\text{I}(\rho_s)^{\text{PT}}\succeq 0\quad \forall \rho_s\in \mathcal{S}_\text{sep}\Bigr\},
    \label{c2}
\end{equation}
where the incapable process matrix \(\rchi_\text{I}\) characterizes  \(\mathcal{E}_\text{I}\), \(\text{PT}\) denotes partial transpose, and \(\mathcal{S}_\text{sep}\) is the set of separable states. Due to linearity and convexity, it suffices to consider pure product states of the form \(\rho_m = \ketbra{m}\), where \(\ket{m} \in \{\ket{0}, \ket{1}, \ket{\pm}, \ket{\pm i}\}^{\otimes 2}\), with \(\ket{\pm} = (\ket{0} \pm \ket{1})/\sqrt{2}\) and \(\ket{\pm i} = (\ket{0} \pm i\ket{1})/\sqrt{2}\). These span the set of separable states relevant for verifying positivity under partial transpose.

Following the procedure for deriving Eq.~\eqref{eq:S15}, the SDP for quantifying the entanglement-creation  composition takes the form:
\begin{equation}
\begin{aligned}
\alpha^\text{cre} = \min_{\Tilde{\rchi}_\text{I}} \quad& 1 - \Tr\bigl(\Tilde{\rchi}_\text{I}\bigr) \\ 
\text{s.t.} \quad 
& \Tilde{\mathcal{E}}_\text{I}(\rho_m)^\text{PT} \succeq 0 \quad \forall m, \\
& \rchi - \Tilde{\rchi}_\text{I} \succeq 0, \quad \Tilde{\rchi}_\text{I} \succeq 0.
\end{aligned}
\label{eq:S26}
\end{equation}

To derive the dual SDP, we introduce Lagrange multipliers for each constraint. This yields the following Lagrangian:
\begin{align}
\mathcal{L} &= 1 - \Tr(\Tilde{\rchi}_\text{I}) 
- \sum_m \Tr\left[ K_m \Tr_1\left( \left( \rho_m^\text{T} \otimes \mathbb{1} \right) \Tilde{\rchi}_\text{I} \right)^\text{PT} \right] 
- \Tr\left[ F(\rchi - \Tilde{\rchi}_\text{I}) \right] 
- \Tr(M \Tilde{\rchi}_\text{I}) \notag \\
&= 1 - \Tr(F \rchi) 
+ \Tr\left[ \left( F - \sum_m \rho_m^\text{T} \otimes K_m^\text{PT} - \mathbb{1} - M \right) \Tilde{\rchi}_\text{I} \right],
\label{eq:S27}
\end{align}
with dual variables \(F \succeq 0\), \(K_m \succeq 0\), and \(M \succeq 0\). The Lagrangian \(\mathcal{L}\) becomes independent of \(\Tilde{\rchi}_\text{I}\) when
\begin{equation}
    M = F - \sum_m \rho_m^\text{T} \otimes K_m^\text{PT} - \mathbb{1}.
\end{equation}
The corresponding dual SDP is:
\begin{equation}
\begin{aligned}
\alpha^\text{cre} = \max_{F, K_m} \quad& 1 - \Tr(F \rchi) \\
\text{s.t.} \quad 
& F - \sum_m \rho_m^\text{T} \otimes K_m^\text{PT} - \mathbb{1} \succeq 0, \\
& F \succeq 0, \quad K_m \succeq 0 \quad \forall m.
\end{aligned}
\label{eq:S28}
\end{equation}

Similarly, the entanglement-creation robustness \(\beta^\text{cre}\), following the procedure for deriving of Eq.~\eqref{eq:S21}, is given by
\begin{equation}
\begin{aligned}
\beta^\text{cre} = \min_{\Tilde{\rchi}_\text{I}} \quad& \Tr(\Tilde{\rchi}_\text{I}) - 1 \\
\text{s.t.} \quad 
& \Tilde{\mathcal{E}}_\text{I}(\rho_m)^\text{PT} \succeq 0 \quad \forall m, \\
& \Tilde{\rchi}_\text{I} - \rchi \succeq 0, \quad \Tilde{\rchi}_\text{I} \succeq 0.
\end{aligned}
\label{eq:S29}
\end{equation}
The associated Lagrangian function is:
\begin{align}
\mathcal{L} &= \Tr(\Tilde{\rchi}_\text{I}) - 1 
- \sum_m \Tr\left[ K_m \Tr_1\left( \left( \rho_m^\text{T} \otimes \mathbb{1} \right) \Tilde{\rchi}_\text{I} \right)^\text{PT} \right]
- \Tr\left[ G(\Tilde{\rchi}_\text{I} - \rchi) \right]
- \Tr(M \Tilde{\rchi}_\text{I}) \notag \\
&= \Tr(G \rchi) - 1 
+ \Tr\left[ \left( \mathbb{1} - G - \sum_m \rho_m^\text{T} \otimes K_m^\text{PT} - M \right) \Tilde{\rchi}_\text{I} \right],
\label{eq:S30}
\end{align}
where \(G \succeq 0\), \(K_m \succeq 0\), and \(M \succeq 0\). The dependence on the primal variable vanishes when
\begin{equation}
    M = \mathbb{1} - G - \sum_m \rho_m^\text{T} \otimes K_m^\text{PT}.
\end{equation}
Substituting into \(\mathcal{L}\) leads to the following dual SDP:
\begin{equation}
\begin{aligned}
\beta^\text{cre} = \max_{G, K_m} \quad& \Tr(G \rchi) - 1 \\
\text{s.t.} \quad 
& \mathbb{1} - G - \sum_m \rho_m^\text{T} \otimes K_m^\text{PT} \succeq 0, \\
& G \succeq 0, \quad K_m \succeq 0 \quad \forall m.
\end{aligned}
\label{eq:S31}
\end{equation}

\begin{table}[t]
 \caption{Calibration data for Qubit 0 and Qubit 1 of the \texttt{ibmq\_belem} device---archived on 11\(^{\text{th}}\) May 2023.}
 \begin{ruledtabular}
 \begin{tabular}{c c c}
 Parameter & Qubit 0  & Qubit 1 \\ \hline
 \(T_1\) (\(\mu\)s) & 106.8263 & 89.1990  \\ 
 \(T_2\) (\(\mu\)s) & 123.8313 & 100.7816  \\ 
 \(prob\_meas0\_prep\_1\) & 0.0398 & 0.0342   \\ 
 \(prob\_meas1\_prep\_0\) & 0.0050 & 0.0080   \\ 
 Readout error & 0.0224 & 0.0211  \\ 
 Readout length (ns) & 6158.2222 & 6158.2222 
 \end{tabular}
 \end{ruledtabular}
 \label{cali}
\end{table}

\begin{figure}[t]
        \centering
        \begin{minipage}[b]{0.48\linewidth}
            \centering
            \includegraphics{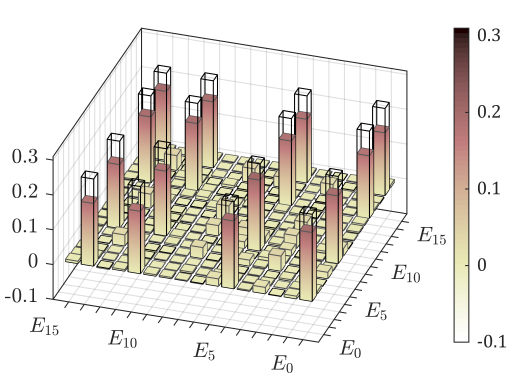}
        \end{minipage}
        \hfill
        \begin{minipage}[b]{0.48\linewidth}
            \centering
            \includegraphics{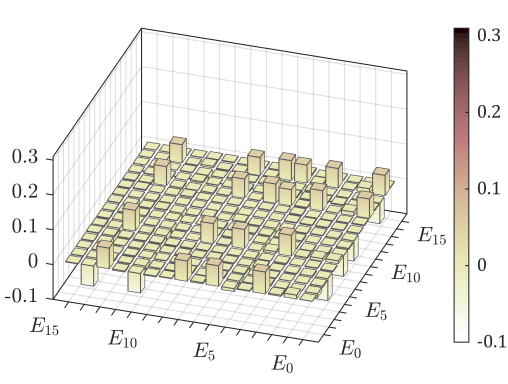}
        \end{minipage}
    \caption{Experimentally reconstructed quantum process matrices \(\rchi_\text{CNOT}\) for the CNOT gate. The real and imaginary components are shown in the left and right, respectively. Here, \(\rchi_\text{CNOT}\) is represented in the operator basis \(\{E_j\}\), where \(E_0=\ketbra{00}\), \(E_1=\ketbra{01}{00}\), \(E_2=\ketbra{10}{00}\), \(\ldots\), and \(E_{15}=\ketbra{11}\). The filled bars indicate the experimentally reconstructed values, while the wireframes represent the corresponding ideal values. The measured process fidelity of the CNOT gate is \(0.7526\).}

    \label{fig:chi_plots}
\end{figure}

\section{Experimental validation: Controlled-NOT gate}

\begin{figure}[t]
        		\centering
		\includegraphics{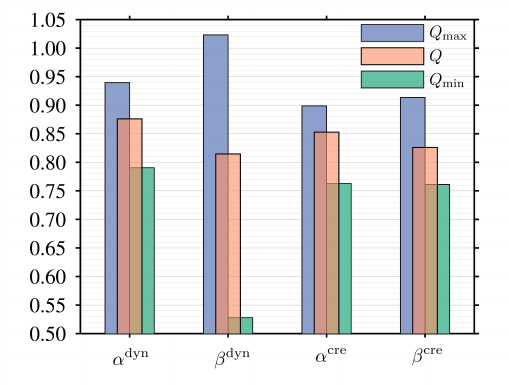}
\caption{
Experimental results of SQPT and AOQPT applied to resource quantification for CNOT gate \(\mathcal{E}_\text{CNOT}\). The measures are shown for both quantum-mechanical dynamics (\(\alpha^{\text{dyn}}, \beta^{\text{dyn}}\)) and entanglement-creation capability (\(\alpha^{\text{cre}}, \beta^{\text{cre}}\)). The actual measures are obtained through the QPC method; the maximum and minimum values are computed through AOQPT.}
    \label{result}
\end{figure}

In this section, we introduce an additional experimental validation of AOQPT by applying it to the characterization of a controlled-NOT (CNOT) gate implemented on an IBM quantum device.

CNOT gate is a two-qubit universal quantum logic gate that plays a central role in gate-based quantum computing. It is fundamental for generating entangled quantum states. When combined with a Hadamard gate, the CNOT enables Bell state preparation and measurement---key primitives for quantum information processing tasks such as quantum teleportation and superdense coding. 

In this work, the CNOT operation is implemented on the 5-qubit \texttt{ibmq\_belem} device with calibration data shown in Table~\ref{cali}. The qubits are superconducting circuits based on Josephson junctions, arranged in a 2D lattice and cooled to millikelvin temperatures~\bluecite{ibmq, qiskit, jjs}. The basis gates include the one-qubit $\mathbb{1}$, $S_X$, $R_Z$, $X$ gates, as well as the two-qubit CNOT gate, operated via precisely controlled microwave pulses with the gate time operation of a few hundred nanoseconds. Quantum states were prepared and measured using 1024 shots without error mitigation. The CNOT gate error rate and duration are 0.0109 and 810.6668 ns, respectively. 

The procedure for quantifying the capabilities of the CNOT operation \(\mathcal{E}_\text{CNOT}\) using the QPC method follows the same approach as for the photonic fusion operation described in the main text. The reconstructed process matrix \(\rchi_\text{CNOT}\) is illustrated in Fig.~\ref{fig:chi_plots}, and the actual values of the measures are presented in Fig.~\ref{result}. The measures indicate that $\mathcal{E}_\text{CNOT}$ exhibits strong quantum-mechanical dynamics and entanglement-generation capabilities, with all values exceeding $0.8$.

As for quantification using AOQPT, the input states \(\{\ket{m_x}\}\) and expected outcomes \(\{\ket{n_y}\}\) are selected from the \(Z \otimes Z\) basis \(\{\ket{00}, \ket{01}, \ket{10}, \ket{11}\}\), while the complementary basis states \(\{\ket{m_u}\}\) and \(\{\ket{n_v}\}\) are chosen from the \(X \otimes X\) basis \(\{\ket{++}, \ket{+-}, \ket{-+}, \ket{--}\}\), with \(\ket{\pm} = (\ket{0} \pm \ket{1})/\sqrt{2}\). As such, this optimization method requires only two experimental settings to construct the stochastic matrices \( (T^{x \rightarrow y}, T^{u \rightarrow v}) \), achieving an approximate \(98\%\) reduction in the number of experimental settings compared to SQPT. Figure~\ref{result} demonstrates that the actual values $Q$ consistently fall within the AOQPT bounds, validating the effectiveness of AOQPT in estimating resource measures \(Q^\text{dyn}\) and \(Q^\text{cre}\). The method provides particularly tight bounds for \(\alpha^\text{dyn}\), \(\alpha^\text{cre}\), and \(\beta^\text{cre}\). Despite the more conservative estimation for \(\beta^\text{dyn}\), the results still highlight the reliability of AOQPT as a resource-efficient and accurate alternative to the conventional approach.